\documentclass[runningheads]{llncs}
\usepackage[T1]{fontenc}
\usepackage{graphicx}
\usepackage{amsmath}
\usepackage[utf8]{inputenc}
\begin{document}
\title{Beyond Individual Influence: The Role of Echo Chambers and Community Seeding in the Multilayer three state $q$-Voter Model}
\titlerunning{Community Seeding in the Multilayer three state $q$-Voter Model}
\author{Igor Hołowacz\and Piotr Bródka\thanks{Corresponding author: \email{piotr.brodka@pwr.edu.pl}}}
\authorrunning{I. Hołowacz et al.}
\institute{Dept. of Artificial Intelligence, Wroc\l{}aw University of Science and Technology, Wroc\l{}aw, Poland; }
\maketitle              % typeset the header of the contribution
\begin{abstract}
The diffusion of complex opinions is severely hindered in multilayer social networks by echo chambers and cognitive consistency mechanisms. We investigate Influence Maximization strategies within the 3-state multilayer $q$-voter model. Utilizing the mABCD benchmark, we simulate social environments ranging from integrated "Open Worlds" to segregated "Fortress Worlds." Our results reveal a topological paradox that we term the "Fortress Trap." In highly modular networks, strategies maximizing local density—such as Clique Influence Maximization (CIM) and k-Shell—fail to trigger global cascades, creating isolated bunkers of consensus due to the "Overkill Effect." Furthermore, we identify a "Redundancy Trap" in perfectly aligned "Clan" topologies, where the structural overlap of layers creates a "Perfect Prison," rendering it the most resistant environment to diffusion. We demonstrate that VoteRank—a strategy that prioritizes diversity of reach over local intensity—consistently outperforms structure-based methods. These findings suggest that, for complex contagion, maximizing topological entropy is more effective than reinforcing local clusters.

\keywords{Opinion Dynamics \and Multilayer Networks  \and q-Voter Model \and Influence Maximization.}
\end{abstract}
\section{Introduction}
\label{sec:intro}

The diffusion of complex opinions requires social reinforcement, a dynamic effectively captured by the multilayer $q$-voter model with the LOCAL \& AND rule~\cite{chmiel2015q}. This framework posits that an agent adopts a new opinion only if they experience unanimous pressure across all social contexts, creating massive systemic inertia. Intuitively, overcoming this resistance should require concentrated reinforcement. Since a single seed is insufficient to flip a node with a threshold of $q=4$, standard logic suggests that "Community Seeding"—targeting dense cliques (CIM~\cite{Venkatakrishna2022CIM}) or cores (k-Shell~\cite{shai2007kshell})—should vastly outperform strategies that target individual dispersed nodes.

In this paper, we challenge this intuition by analyzing seeding efficiency on realistic topologies generated by the mABCD benchmark~\cite{krainski2025multilayer}. Our results reveal a counter-intuitive phenomenon we term the "Fortress Trap". We demonstrate that in highly modular networks, strategies aimed at maximizing local density (CIM, k-Shell) paradoxically fail, creating isolated "bunkers" of consensus that cannot propagate. Although standard Hub Seeding (Degree) remains a robust baseline, the most surprising efficacy is exhibited by \texttt{VoteRank}~\cite{zhang2016identifying} and \texttt{PageRank}~\cite{page1999pagerank}. Despite distributing seeds sparsely rather than concentrating them, these centrality-based methods outperform structural strategies, suggesting that in fragmented digital landscapes, the diversity of reach is more critical than local intensity for triggering global cascades.

The remainder of this paper is organized as follows. Section \ref{sec:related} outlines the theoretical foundations of opinion dynamics and multilayer networks. Section \ref{sec:model} formally defines the microscopic rules of the 3-state multilayer $q$-voter model with the LOCAL \& AND update mechanism. In Section \ref{sec:mfa}, we derive the mean-field approximation to understand the global phase space dynamics. Section \ref{sec:experimental_setup} details the simulation protocol and the generation of diverse social topologies using the mABCD benchmark. Section \ref{sec:results} presents our main empirical findings, demonstrating the "Fortress Trap" while evaluating the performance of specific seeding strategies. Finally, Section \ref{sec:conclusion} summarize the implications of topological alignment on diffusion and outline directions for future research.

\section{Related Work}
\label{sec:related}

The study of how opinions, behaviors, and innovations propagate through society lies at the intersection of statistical physics, computer science, and sociology. Our work builds upon three fundamental pillars of this interdisciplinary field: non-linear opinion dynamics, multilayer network theory, and the algorithmic challenge of influence maximization.

\subsection{Opinion Dynamics and Conformity Models}
Traditional approaches to information diffusion, such as the Independent Cascade or Linear Threshold models, often treat humans as passive nodes in a contagion process. However, adopting a political opinion or a complex innovation requires active social reinforcement. The application of statistical physics to social dynamics—popularized as sociophysics and comprehensively reviewed by Castellano et al.~\cite{castellano2009}—introduced Ising-like models to describe this phenomenon, notably initiated by the Sznajd model~\cite{sznajd2000opinion}. 
Building upon these foundations, Castellano, Mu{\~n}oz, and Pastor-Satorras introduced the non-linear $q$-voter model~\cite{castellano2009nonlinear}, which has since become a cornerstone of this domain. It captures the essence of social conformity, where a target agent adopts a new opinion only if influenced by a unanimous group of $q$ neighbors. Recent extensions of this framework have gone beyond binary choices by incorporating a third "undecided" state~\cite{vieira2020phase}. This crucial addition prevents artificial polarization and captures the hesitation that characterizes real-world decision-making processes.

\subsection{Multilayer Networks and Cognitive Consistency}
Early network science primarily focused on monoplex topologies. However, as comprehensively summarized by Kivelä et al.~\cite{kivela2014multilayer}, human interaction is inherently multiplex: individuals communicate via distinct channels (e.g., family, work, online media), each with its own topology. 

Research into multilayer opinion dynamics, such as the work by Jędrzejewski et al.~\cite{jedrzejewski2017phase}, revealed that multiplexity dramatically alters phase transitions. Chmiel and Sznajd-Weron~\cite{chmiel2015q} introduced the LOCAL \& AND aggregation rule. This mechanism posits that an agent requires consistent reinforcement across all layers to change their state, effectively modeling cognitive consistency—the psychological resistance to conflicting signals from different environments.

\subsection{Influence Maximization and Echo Chambers}
Finding the optimal set of seed nodes to trigger a global cascade is the fundamental problem of Influence Maximization (IM), While traditional strategies target hubs, real-world networks are fragmented into "echo chambers" that resist external influence~\cite{del2016spreading}. In the context of complex contagion ($q=4$), this creates a strategic dilemma: does overcoming inertia require concentrating resources within dense communities (via structural strategies like CIM), or distributing them to bridge isolated clusters (via centrality measures like VoteRank)? To resolve this, we employ the mABCD benchmark, which allows us to systematically tune the insularity of social bubbles ($\xi$) and rigorously test whether "bunker" or "bridge" strategies are superior in breaking multilayer inertia.

\section{The Multilayer $q$-Voter Model}
\label{sec:model}

The clasic $q$-voter model~\cite{castellano2009nonlinear}, operates on a binary state space and describes the non-linear nature of social conformity. We extend this framework to a 3-state system evolving on a multilayer topology, which captures both the multi-contextual nature of human interactions and the psychological state of hesitation.

Let the social structure be represented as a multiplex network $\mathcal{M} = (V, \vec{E})$, consisting of a fixed set of $N$ nodes (agents) $V = \{1, 2, \dots, N\}$ and a vector of edge sets $\vec{E} = \{E_1, E_2, \dots, E_L\}$, one for each layer $\alpha \in \{1, \dots, L\}$. Each layer is described by adjacency matrix $A^{[\alpha]}$, where $A_{ij}^{[\alpha]} = 1$ if agents $i$ and $j$ are connected in layer $\alpha$, and $0$ otherwise. 
At any discrete time step $t$, the state of the system is defined by the configuration vector $\vec{\sigma}(t) = [\sigma_1(t), \sigma_2(t), \dots, \sigma_N(t)]$. Unlike traditional Ising-like models, we allow agents to occupy one of three distinct states, $\sigma_i(t) \in \{+1, -1, 0\}$, which we denote as $A$, $B$, and $U$ (Undecided), respectively. The state $U$ represents agents who are currently unexposed, indifferent, or actively in doubt. This addition prevents the system from simplifying to a trivial two-party tug-of-war and accurately reflects real-world diffusion processes where a substantial portion of the population remains neutral.

The evolution of the system is modeled as a continuous-time Markov process, simulated via random sequential updating. Time is measured in Monte Carlo Steps (MCS), where one MCS corresponds to $N$ elementary update events. In each elementary step, a single target agent $i$ is chosen uniformly at random.

The behavior of the agent is governed by two complementary mechanisms: independence (with probability $p$) and conformity (with probability $1-p$). This duality is central to sociophysics, reflecting the constant interplay between individual free will and social pressure.

\subsection{The Independence Mechanism (Noise)}
With probability $p$, agent $i$ disregards the network topology and acts autonomously. We assume that individuals with an established opinion ($A$ or $B$) are subject to "opinion decay" or memory loss, moving to the undecided state. Conversely, undecided agents may spontaneously adopt an opinion through random discovery. The transition probabilities under independence are defined as:
\begin{align*}
    P(\sigma_i(t+\Delta t) = U \mid \sigma_i(t) \in \{A, B\}) &= \frac{1}{2} \\
    P(\sigma_i(t+\Delta t) = C \mid \sigma_i(t) = U) &= \frac{1}{3} \quad \text{for } C \in \{A, B, U\}
\end{align*}

\subsection{The Conformity Mechanism (Multilayer $q$-influence)}
With probability $1-p$, the agent is susceptible to social influence. We apply the LOCAL \& AND aggregation rule~\cite{chmiel2015q}, which assumes that individuals require consistent reinforcement across all social contexts (layers) to change their minds. 

For each layer $\alpha$, the target agent samples a subset (panel) of $q$ neighbors, denoted as $\mathcal{S}_i^{[\alpha]}$. Crucially, the sampling is performed \textit{with replacement}. This implies that an agent $i$ with degree $k_i^{[\alpha]} < q$ can still be influenced, as neighbors are consulted multiple times (simulating repeated interactions or high-frequency exposure). An agent with degree $k_i^{[\alpha]} = 0$ experiences no pressure in that layer.

Influence occurs if and only if two restrictive conditions are met:
\begin{enumerate}
    \item \textbf{Intra-layer unanimity:} Within each layer $\alpha$, all $q$ sampled neighbors hold the same opinion $C_{\alpha} \in \{A, B\}$. Formally: $\forall_{j \in \mathcal{S}_i^{[\alpha]}} \sigma_j(t) = C_{\alpha}$.
    \item \textbf{Inter-layer agreement:} The unanimous opinion is identical across all $L$ layers. Formally: $C_1 = C_2 = \dots = C_L = C$.
\end{enumerate}

\begin{figure}[htbp]
    \centering   \includegraphics[width=0.80\textwidth]{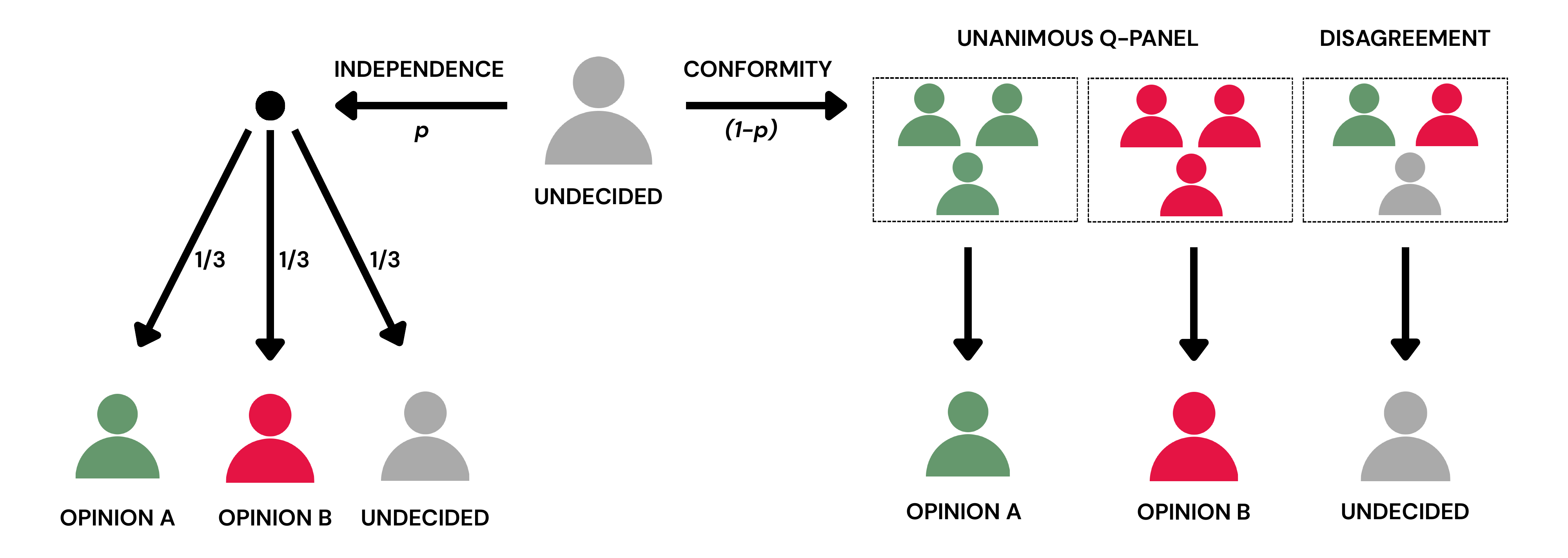}
    \caption{Transitions for an Undecided agent ($U$). Conformity with probability $1-p$ (requiring a $q$-sized unanimous panel across all $L$ layers). Spontaneous transitions driven by independence (noise with probability $p$).}
    \label{fig:trans_u}
\end{figure}

\begin{figure}[htbp]
    \centering
    \includegraphics[width=0.80\textwidth]{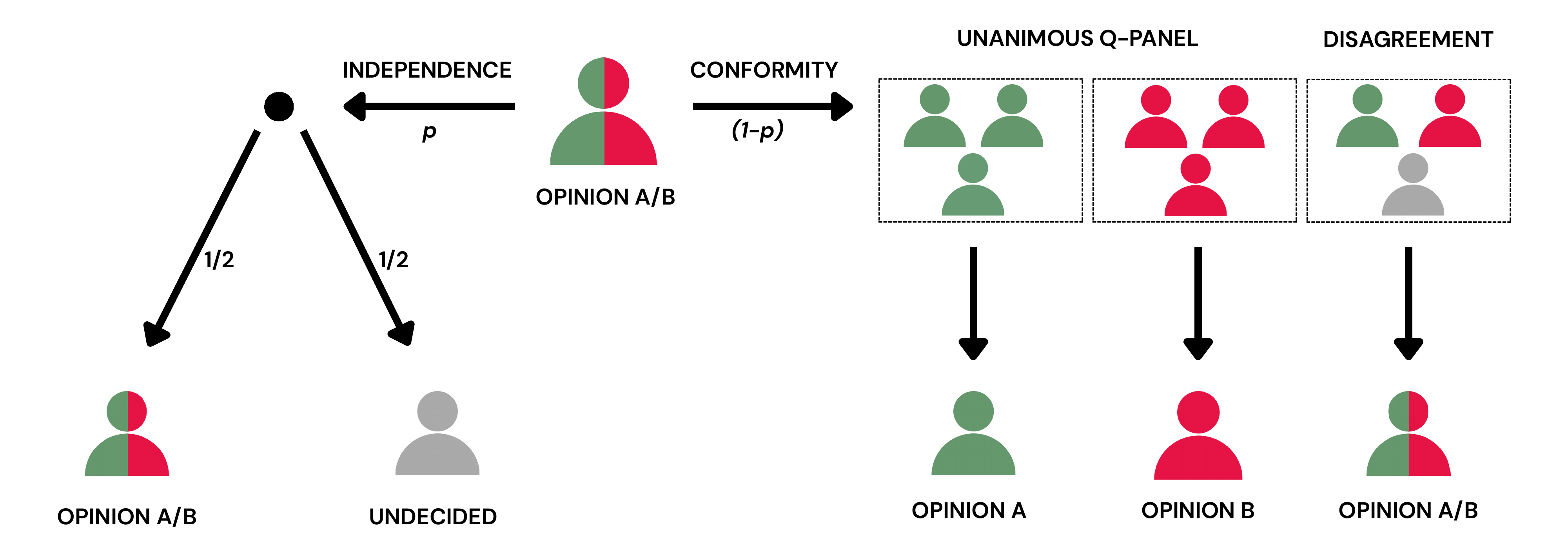}
    \caption{Transitions for a Decided agent ($A$ or $B$).}
    \label{fig:trans_d}
\end{figure}

If the global consensus $C$ exists and differs from the current state $\sigma_i(t)$ (i.e., the target is undecided or holds the opposing view), the agent adopts opinion $C$. Otherwise, no state change occurs. This strict condition mirrors the psychological principle of cognitive consistency—people resist adopting an opinion if they receive conflicting signals from different environments (e.g., family vs. workplace).
To provide a clear visual summary of these dynamics, Figures \ref{fig:trans_u} and \ref{fig:trans_d} illustrate the possible state transitions for undecided and decided agents, respectively. 
As shown in Figure \ref{fig:trans_u}, an undecided agent can transition to a decided state ($A$ or $B$) via two distinct pathways: either spontaneously through the independence mechanism (noise) or by yielding to social conformity when exposed to a unanimous panel. Conversely, Figure \ref{fig:trans_d} highlights the distinct resilience of decided agents. The independence mechanism acts exclusively as an "opinion decay", spontaneously pushing decided agents back into the undecided state. A direct transition to the opposing view (e.g., from $A$ to $B$) is impossible through noise alone; it strictly requires an overwhelming, coordinated social pressure from a unanimous neighborhood across all layers.


\section{Mean-Field Analysis and Topological Validation}
\label{sec:mfa}

To understand the global constraints of this microscopic rule, we first analyze the system in the Mean-Field Approximation (MFA) limit. This corresponds to an infinite, fully mixed population (Complete Graph, CG) where the network topology is ignored, and interactions are purely probabilistic.

\subsection{Mean-Field Formulation and Phase Transitions}
\label{sec:4_1}

Let $c_A$, $c_B$, and $c_U$ denote the global concentrations of agents in states $A$, $B$, and $U$, respectively ($c_A + c_B + c_U = 1$). Under the LOCAL \& AND rule, the probability of assembling a unanimous panel of size $q$ across $L$ statistically independent layers scales as $(c_A^q)^L = c_A^{qL}$. The time evolution of the system is governed by the following coupled ordinary differential equations:

\begin{align}
    \frac{dc_A}{dt} &= c_U \left( \frac{p}{3} + (1-p)c_A^{qL} \right) + c_B(1-p)c_A^{qL} - c_A\frac{p}{2} - c_A(1-p)c_B^{qL} \label{eq:mfa_A} \\
    \frac{dc_B}{dt} &= c_U \left( \frac{p}{3} + (1-p)c_B^{qL} \right) + c_A(1-p)c_B^{qL} - c_B\frac{p}{2} - c_B(1-p)c_A^{qL} \label{eq:mfa_B}
\end{align}

\begin{figure}[htbp]
    \centering
    \includegraphics[width=0.6\textwidth]{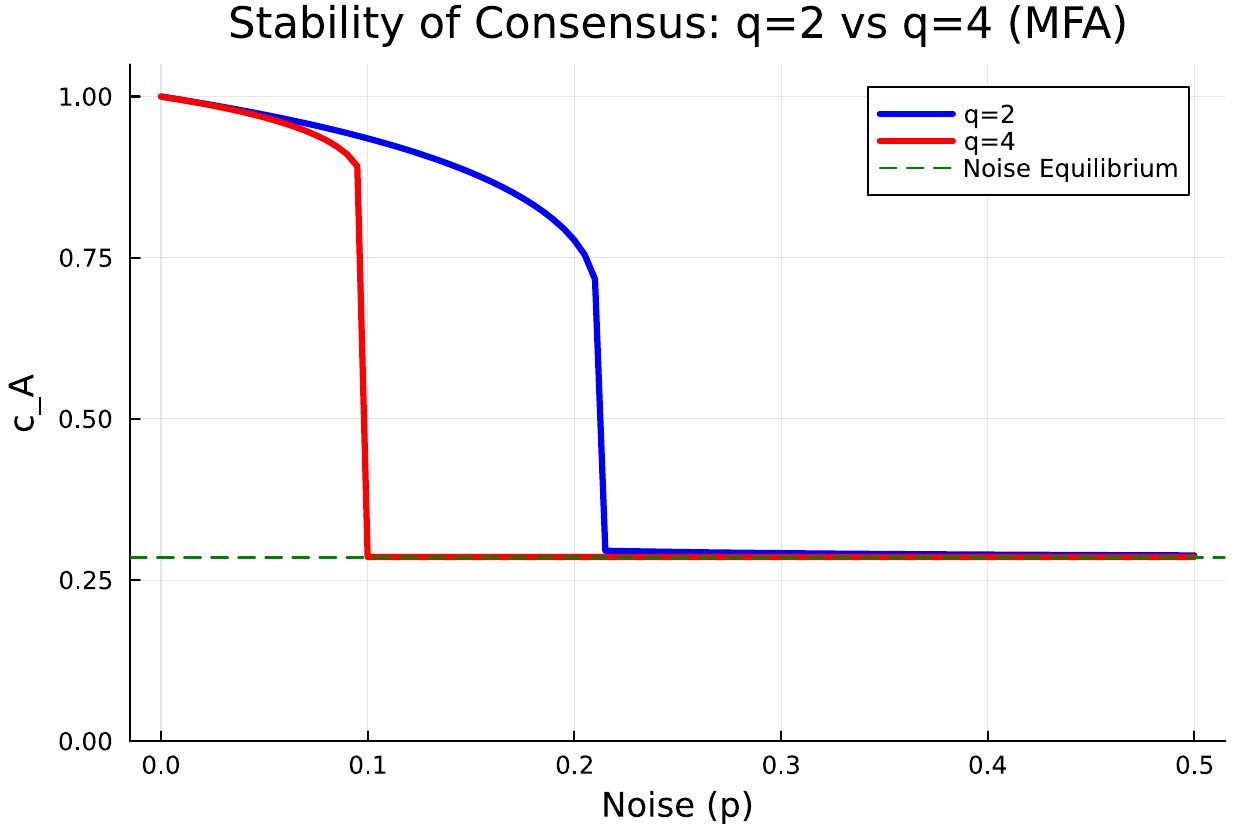}
    \caption{Mean Field Approximation results for $q=2$ vs. $q=4$ ($L=2$). The $q=2$ case allows for a wider range of stability with gradual degradation, while $q=4$ maintains higher consensus quality but collapses at a lower noise threshold.}
    \label{fig:mfa_regimes}
\end{figure}

Figure \ref{fig:mfa_regimes} compares the phase transitions for moderate ($q=2$) and high ($q=4$) influence thresholds in a multiplex system ($L=2$). 
The $q=2$ regime exhibits higher stability, sustaining a majority consensus up to a critical noise level of $p_c > 0.2$. However, the consensus level declines gradually as noise increases.
In contrast, the $q=4$ regime is characterized by a lower critical threshold ($p_c \approx 0.1$), indicating reduced stability against noise. Despite this narrower stability range, the system maintains a near-perfect consensus ($c_A \approx 1.0$) until the critical point, where it undergoes a sharp, discontinuous transition to the noise equilibrium.

\subsection{Phase-Space Dynamics and Basins of Attraction}
To visualize the global implications of these dynamics, we project the system's states onto a 2D simplex defined by $(c_A, c_B)$. This phase portrait allows us to trace the deterministic trajectories from a grid of various initial configurations.

\begin{figure}[htbp]
    \centering
    \includegraphics[width=0.8\textwidth]{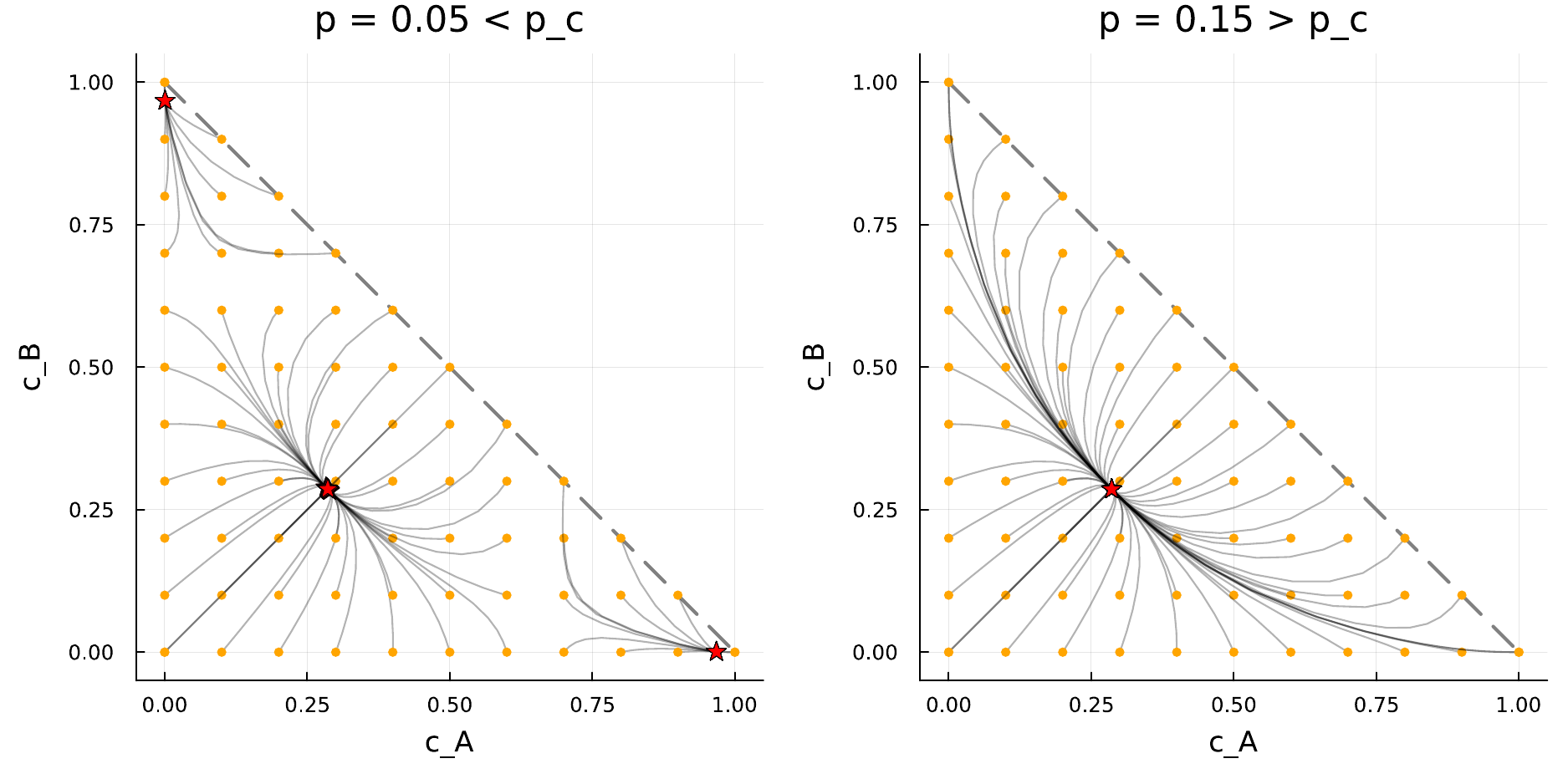}
    \caption{Deterministic phase-space trajectories for the multilayer $q$-voter model ($q=4, L=2$). Orange dots indicate starting configurations, and red stars indicate final stationary states. Left: For low noise ($p=0.07$), multiple attractors exist depending on initial conditions. Right: For high noise ($p=0.15$), all trajectories collapse into the central Noise Equilibrium.}
    \label{fig:phase_space_mfa}
\end{figure}

Figure \ref{fig:phase_space_mfa} compares the phase portraits for two distinct noise regimes. The left panel ($p=0.05$) illustrates the dynamics within the bistable region ($p < p_c$). Here, the system is dominated by historical inertia. A strong initial bias toward opinion $A$ or $B$ successfully overcomes the noise, driving the system toward the respective extreme attractors. Conversely, regions of high indecision or symmetrical split (the center of the simplex) fall into the Basin of Attraction of the central Noise Equilibrium.
The right panel ($p=0.15$) dramatically demonstrates the total collapse of systemic memory. As the noise exceeds the critical threshold, the multiple distinct attractors vanish. Irrespective of the initial conditions—even starting from perfect consensus ($c_A=1$)—all trajectories are inexorably pulled into a single, global point-attractor: the Noise Equilibrium.

\section{Experimental Design and Parameter Calibration}
\label{sec:experimental_setup}

To systematically evaluate the model, we established a rigorous simulation protocol based on the \textbf{mABCD} (Multilayer Artificial Benchmark for Community Detection) model. This generative framework allows for the construction of controlled social topologies where we can independently tune community structure, hierarchy, and interlayer correlations.

\subsection{Network Environment}

We simulate a population of $N=1,000$ agents interacting within a duplex network structure ($L=2$). The layers represent distinct social contexts (e.g., professional vs. private sphere).

\subsubsection{Fixed Micro-Structure (The Baseline)}
The network micro-structure is constrained by fixed parameters defining a society of small, tight-knit groups without global dominators:
\begin{itemize}
    \item \textbf{Degree Distribution ($\gamma=2.5, \delta=2, \Delta=25$):} Degrees follow a power law bounded strictly by $\Delta=25$. This prevents the formation of "Super-Hubs," forcing reliance on local leaders.
    \item \textbf{Community Structure ($\beta=1.5, s \in [16, 25]$):} The restricted community size ($s \approx 20$) reflects the structure of close-knit cliques or functional teams rather than broad social collectives.
\end{itemize}

\subsubsection{Variable Parameters (Sensitivity Analysis)}
To determine seeding boundary conditions, we systematically vary three macroscopic parameters:
\begin{enumerate}
    \item \textbf{Community Isolation ($\xi$):} Varies from $\mathbf{0.05}$ (Strong Segregation, 95\% internal edges) to $\mathbf{0.35}$ (Open Society).
    \item \textbf{Partition Correlation ($r$):} Determines group overlap. Varies from $\mathbf{0.0}$ (Disjoint Contexts, randomized groups) to $\mathbf{1.0}$ (Aligned Contexts, identical groups across layers).
    \item \textbf{Degree Correlation ($\rho$):} Determines hierarchy consistency between layers. Varies from $\mathbf{0.0}$ (Fragmented Influence) to $\mathbf{0.9}$ (Consolidated Elites, where Hubs overlap).
\end{enumerate}

%\begin{table}[htbp]
%\centering
%\caption{Summary of model parameters. Fixed parameters define the micro-structure, while variable parameters explore the topological phase space.}
%\label{tab:params}
%\begin{tabular}{@{}llcc@{}}
%\textbf{Symbol} & \textbf{Parameter} & \textbf{Value}\\
%\hline
%$N, L$ & Network Size & $1000$, $2$ layers \\
%$\gamma$ & Degree Exponent & $2.5$ \\
%$\delta, \Delta$ & Degree Bounds & $2, 25$ \\
%$\beta$ & Community Size Exponent & $1.5$ \\
%$s, S$ & Community Size Range & $16, 25$ \\
%$q$ & Active actors & $1$ %\\

%\end{tabular}
%\end{table}
\subsection{Investigated Social Worlds}
To evaluate the robustness of seeding strategies under diverse sociological conditions, we constructed a grid of six distinct network scenarios. These scenarios are generated by systematically varying three macroscopic control parameters while maintaining the fixed micro-structure defined in the previous section.

\subsubsection{Variable Parameters}
We manipulate the network topology along two principal dimensions:

\begin{enumerate}
    \item \textbf{Community Isolation ($\xi$):} This parameter controls the permeability of information boundaries between social groups.
    \begin{itemize}
        \item \textbf{High Segregation ($\xi = 0.05$):} Models a "Fortress" society where 95\% of interactions occur within local micro-communities. This creates strong echo chambers and limits global diffusion.
        \item \textbf{High Mixing ($\xi = 0.35$):} Models an "Open Society". With 35\% of edges connecting to external groups, the social structure is fluid, facilitating rapid information flow but weakening local peer pressure.
    \end{itemize}

    \item \textbf{Inter-layer Correlations ($\rho, r$):} These parameters define the consistency of social structures across layers. We distinguish three archetypes:
    \begin{itemize}
        \item \textbf{Chaos ($\rho \approx 0.1, r=0$):} Represents disjoint social contexts with inconsistent hierarchy. An influential node in Layer 1 is likely insignificant in Layer 2, and community memberships are randomized.
        \item \textbf{Elite ($\rho \approx 0.9, r=0$):} Represents a society with consolidated status but disjoint groups. Social circles differ between layers, but the same "Elite" individuals maintain high degree centrality in both contexts.
        \item \textbf{Clan ($\rho \approx 0.9, r=1$):} Represents a fully integrated "Small Town" dynamic. Not only is social status preserved, but the community structure is identical across layers ($C_1 \equiv C_2$). This creates a reinforcement effect, where agents interact with the same peers in both private and professional spheres.
    \end{itemize}
\end{enumerate}

\subsubsection{Scenario Grid}
By combining the two levels of isolation with the three types of correlation, we obtain a set of six experimental environments. The values of the specific parameters for each scenario are summarized in Table \ref{tab:parameters}.

\begin{table}[h!]
\centering
\caption{Parameter values for the six investigated network scenarios. $\xi$ denotes the mixing parameter, $\rho$ the inter-layer degree correlation, and $r$ the partition correlation.}
\label{tab:parameters}
\begin{tabular}{lccc}
\hline
\textbf{Scenario Name} & \textbf{Level of noise ($\xi$)} & \textbf{Status ($\rho$)} & \textbf{Groups ($r$)} \\ 
\hline
\multicolumn{4}{l}{\textit{Fortress Worlds (High Segregation)}} \\
1. \texttt{Fortress: Chaos} & 0.05 & $ 0.1$ & 0.0 \\
2. \texttt{Fortress: Elite} & 0.05 & $ 0.9$ & 0.0 \\
3. \texttt{Fortress: Clan}  & 0.05 & $ 0.9$ & 1.0 \\ 
\hline
\multicolumn{4}{l}{\textit{Open Worlds (High Mixing)}} \\
4. \texttt{Open: Chaos}     & 0.35 & $ 0.1$ & 0.0 \\
5. \texttt{Open: Elite}     & 0.35 & $ 0.9$ & 0.0 \\
6. \texttt{Open: Clan}      & 0.35 & $ 0.9$ & 1.0 \\ 
\hline
\end{tabular}
\end{table}

\section{Experimental Results}\label{sec:results}

\subsection{Phase I: Baseline Stability Analysis}
\begin{figure}[h!]
    \centering
    \includegraphics[width=0.6\textwidth]{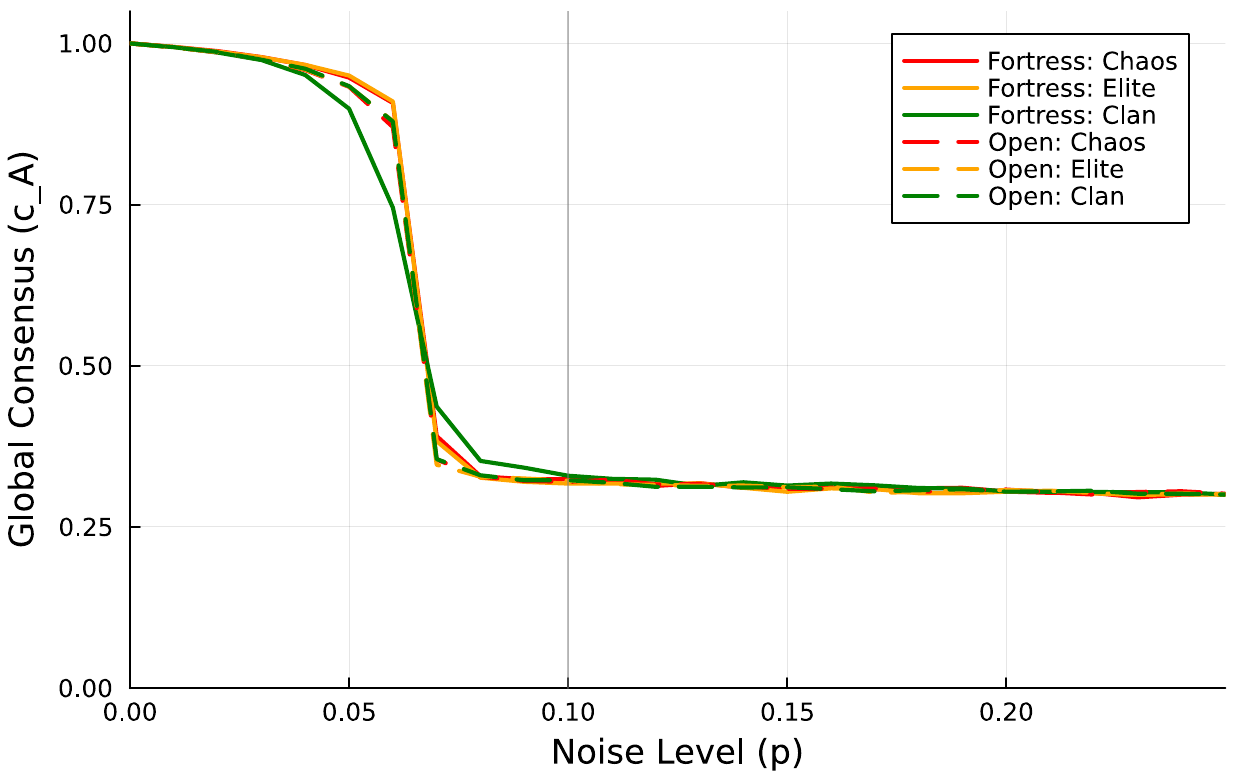}
    \caption{Phase transition diagrams ($q=4$). %The critical point $p_c$ marks the transition to disorder.
    }
    \label{fig:phase_diagram}
\end{figure}

We first examine the spontaneous breakdown of consensus from an initially ordered state ($c_A(0)=1$) under varying noise levels $p$. Figure \ref{fig:phase_diagram} presents the order parameter $c_A$ averaged over $100$ independent realizations after $1000$ MCS.

A phase transition is observed in the interval $p_c \in [0.06, 0.07]$. This threshold is significantly lower than Mean Field predictions, a discrepancy attributed to the \textit{finite connectivity effect}. In sparse networks, assembling a unanimous neighborhood of size $q=4$ is topologically constrained, making the system susceptible even to low noise levels. The \texttt{Fortress Clan} topology exhibits a slightly more gradual decay, as high modularity confines consensus within local clusters.

\subsection{Phase II: Seeding Effectiveness and Temporal Dynamics}
In the second phase, we evaluate the ability of selected strategies to drive consensus from a globally undecided state ($S_i=0$). The simulations cover three budgets $f \in \{0.03, 0.05, 0.10\}$ and three noise regimes: $p=0.0$ (noise-free structural baseline), $p=0.03$ (metastable) and $p=0.06$ (critical). Following~\cite{czuba2025rankrefiningseedselectionmethods}, we compare six strategies: \texttt{Random}, \texttt{Degree}, \texttt{PageRank}, \texttt{VoteRank}, \texttt{k-Shell}, and \texttt{CIM}.

\begin{figure}[h!]
    \centering
    \includegraphics[width=0.95\textwidth]{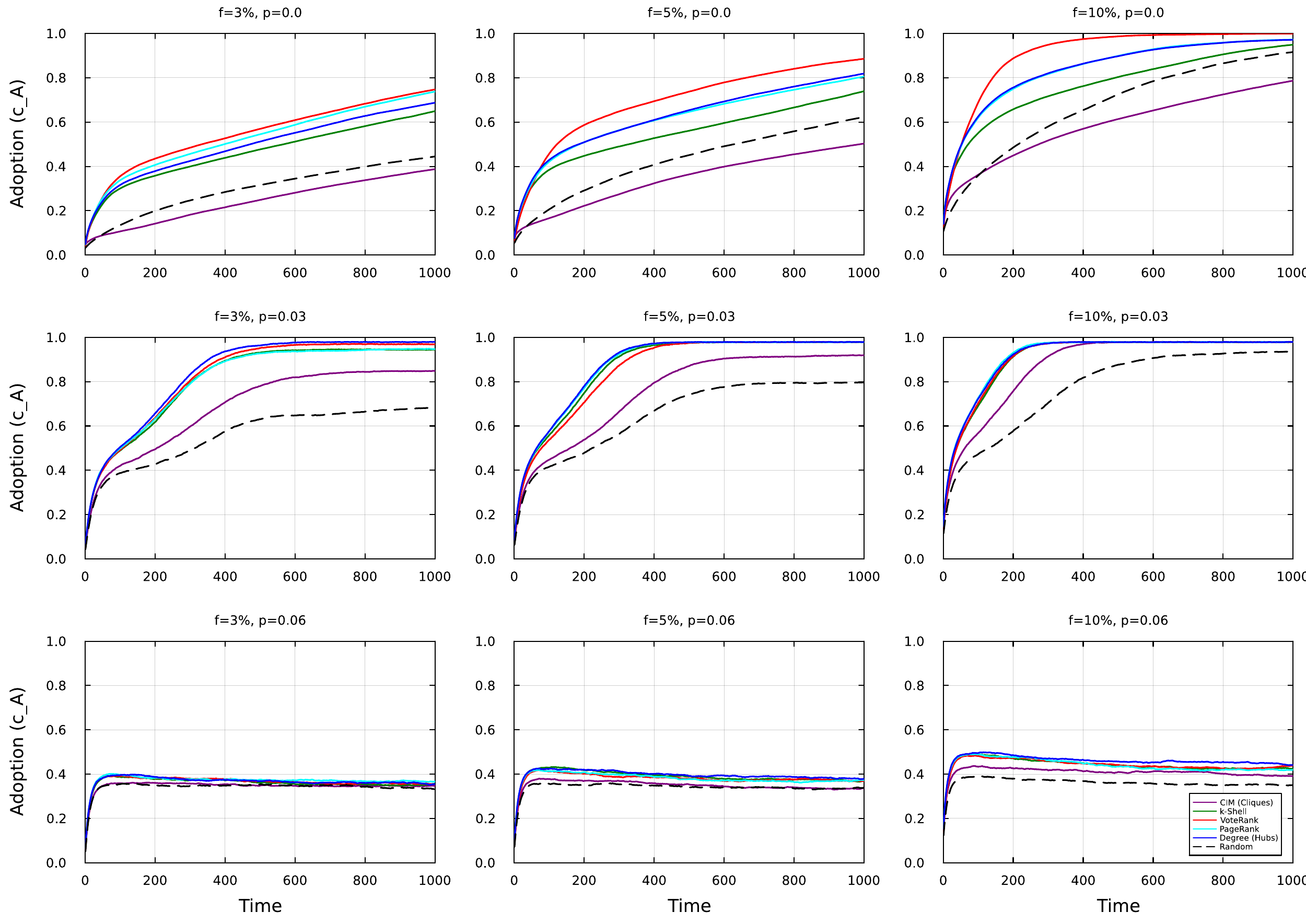}
    \caption{\texttt{Fortress: Chaos}. Strategies: \texttt{VoteRank} (red), \texttt{PageRank} (cyan), \texttt{Degree} (blue), \texttt{k-Shell} (green), \texttt{CIM} (purple), \texttt{Random} (black dashed). \texttt{VoteRank} dominates at $p=0$. \texttt{CIM} strategie perform poorly, falling near or below the random baseline.}
    \label{fig:traj_fortress_chaos}
\end{figure}

\begin{figure}[h!]
    \centering
    \includegraphics[width=0.95\textwidth]{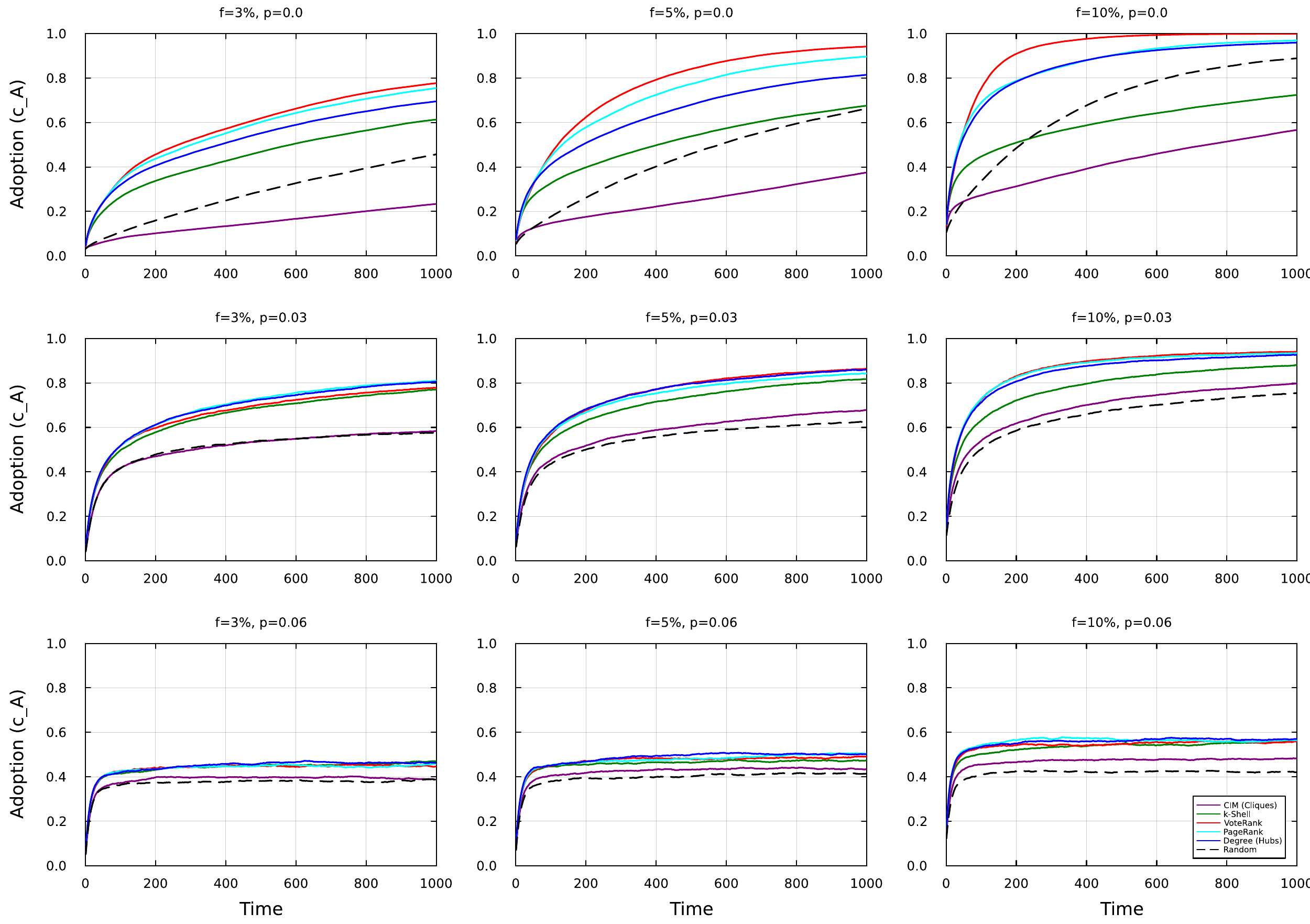}
    \caption{\texttt{Fortress: Clan}. Strategies: \texttt{VoteRank} (red), \texttt{PageRank} (cyan), \texttt{Degree} (blue), \texttt{k-Shell} (green), \texttt{CIM} (purple), \texttt{Random} (black dashed). Despite layer alignment, \texttt{CIM} remains the least effective strategy. \texttt{VoteRank} maintains its lead at $p=0$.}
    \label{fig:traj_fortress_clan}
\end{figure}

\begin{figure}[h!]
    \centering
    \includegraphics[width=0.95\textwidth]{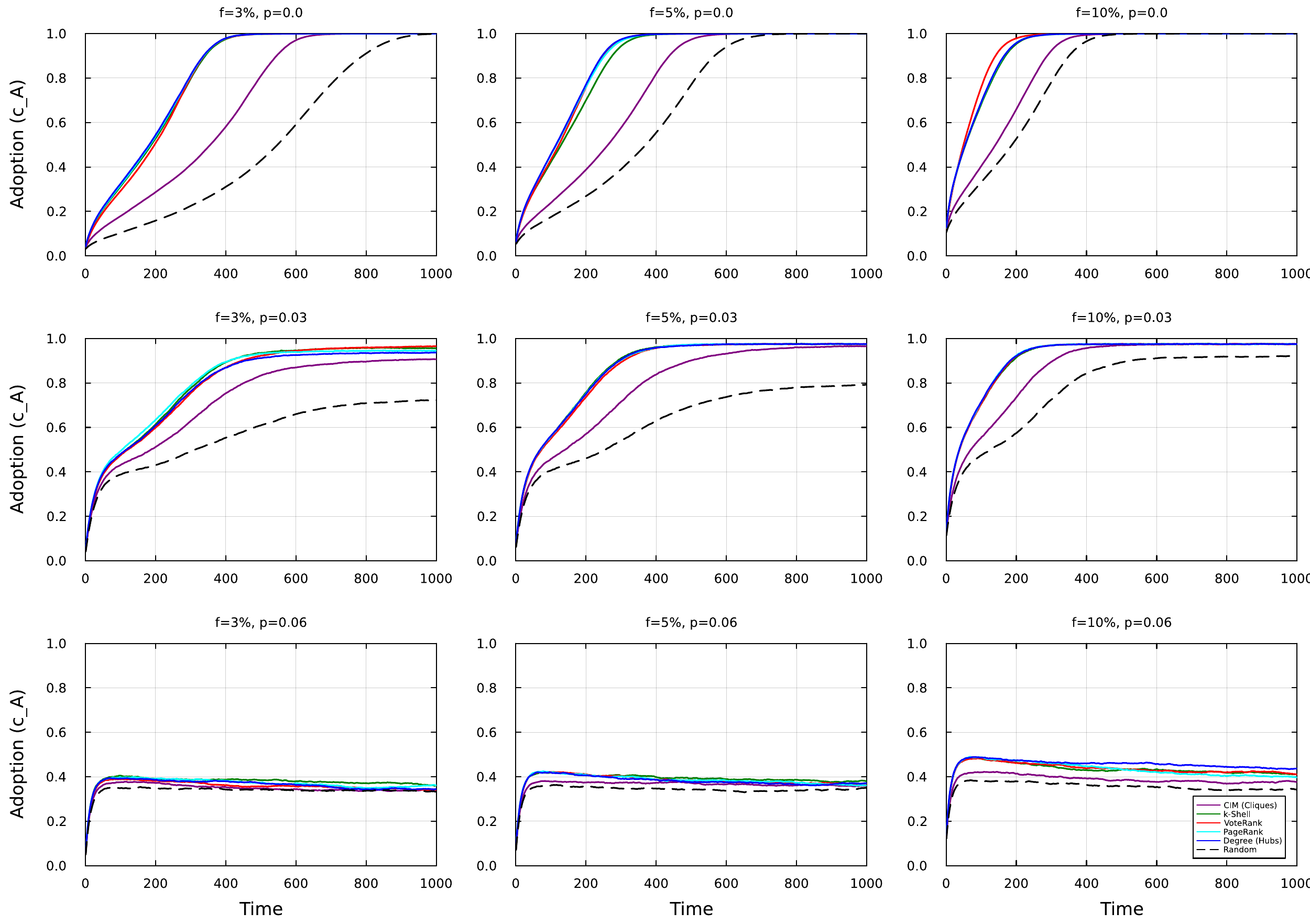}
    \caption{\texttt{Open: Clan}. Strategies: \texttt{VoteRank} (red), \texttt{PageRank} (cyan), \texttt{Degree} (blue), \texttt{k-Shell} (green), \texttt{CIM} (purple), \texttt{Random} (black dashed). High mixing ($\xi=0.35$) compresses the performance gap.}
    \label{fig:traj_open_clan}
\end{figure}

\subsubsection{Analysis of Results}
The experimental results contradict the initial intuition regarding the advantage of dense structures in complex contagion ($q=4$). A clear hierarchy emerges, driven by the trade-off between seed diversity and local redundancy.

\paragraph{The Failure of Dense Structures (CIM \& k-Shell).}
Across all investigated scenarios, \texttt{CIM} (purple) consistently performs the worst, failing to initiate significant diffusion. Similarly, \texttt{k-Shell} (green) struggles, particularly in \texttt{Fortress} environments, where its performance drops near or even below the \texttt{Random} baseline (black dashed). This indicates a severe "Overkill Effect": in a sparse network requiring $q=4$ active neighbors, concentrating seeds in cliques or cores creates isolated "bunkers" that are internally stable but topologically disconnected from the rest of the system.

\paragraph{Hierarchy at Noiseless Baseline ($p=0$).}
In the absence of noise, a distinct order of efficiency is observed, particularly for low budgets ($f=3\%, 5\%$): \textbf{VoteRank~(Red):} By explicitly penalizing the neighbors of selected nodes, this strategy maximizes the spatial distribution of seeds. It avoids "preaching to the choir," which is the most effective approach in fragmented Fortress worlds.
\textbf{PageRank~(Cyan):} As a measure of global centrality, it identifies bridges between clusters, outperforming local metrics.

\paragraph{The Paradoxical Role of Noise.}
Comparing the baseline ($p=0.0$) with the metastable regime ($p=0.03$), we observe an inversion effect. Stochastic noise appears to degrade the performance of "precision" strategies like \texttt{VoteRank} and \texttt{PageRank}, whose calculated advantage relies on the deterministic structure of the network. Conversely, noise facilitates the "brute force" strategies—\texttt{Degree}, \texttt{k-Shell}, and \texttt{CIM}. For these methods, random flipping likely helps break local deadlocks (metastable traps) caused by their excessive concentration of seeds, allowing the infection to escape the initial clusters.

\section{Conclusions and Future Work}
\label{sec:conclusion}

In this work, we challenged the prevailing assumption that identifying and targeting dense community structures is the universal key to solving the Influence Maximization problem. By systematically scanning the phase space of the 3-state multilayer $q$-voter model, we demonstrated that the efficacy of seeding strategies is strictly conditional on the topological alignment of social layers.

Our most significant finding is the identification of the "Fortress Trap". In highly modular networks ("Fortress Worlds"), strategies that maximize local reinforcement—such as Clique Influence Maximization (CIM) and k-Shell decomposition—become counter-productive. For complex contagion ($q=4$), these methods create "bunkers" of consensus that are internally stable but topologically isolated. Instead of triggering a cascade, the seed energy is dissipated within closed loops due to the "Overkill Effect."

Furthermore, our results regarding the Fortress: Clan scenario reveal a "Redundancy Trap". Contrary to the hypothesis that inter-layer alignment ($r=1$) acts as an incubator for spreading, we found that perfect alignment creates a "Perfect Prison." The structural redundancy of overlapping layers reduces the effective reach of seeds—bridges in Layer 1 lead to the same communities as bridges in Layer 2. Consequently, the lack of topological disorder inhibits the "leakage" of opinions to new clusters.
In such fragmented environments, the \texttt{VoteRank} strategy is superior. By penalizing redundancy and forcing a wide distribution of seeds, it effectively "mines" multiple communities simultaneously, overcoming the systemic inertia caused by modularity. Ultimately, our results suggest a shift in the philosophy of influence operations: in a chaotic, fragmented digital landscape, diversity of reach (entropy maximization) is more valuable than local intensity (clustering maximization).

%\section{Future Work}
%\label{sec:future}

Presented framework opens several promising avenues for further research:
\textbf{Heterogeneous Thresholds.} We assumed a uniform conformity threshold ($q=4$) for all agents. In reality, psychological resistance varies. Future studies should incorporate heterogeneous thresholds ($q_i$, potentially correlated with the degree of the node or the position in the community.
\textbf{Real-World Multiplex Validation.} While the mABCD benchmark provides a controlled testing ground, validating these findings on large-scale empirical datasets—such as the intersection of Twitter (political layer) and Foursquare (physical layer)—is a necessary next step. We hypothesize that real social networks lie somewhere between the "Chaos" and "Clan" extremes, requiring hybrid seeding strategies.
\textbf{Competitive Influence.} Our model assumes a single active opinion fighting against neutrality ($S_i=0$). A natural extension is to introduce a competitive setting (Red Seeds vs. Blue Seeds), where two opposing views fight for dominance. Understanding whether "Bridge" strategies (VoteRank) or "Bunker" strategies (CIM) prevail in a zero-sum conflict would have profound implications for political campaign modeling.
\textbf{Temporal Networks.} Finally, social ties are not static. Introducing temporal dynamics—where edges flicker or communities evolve over time—could reveal whether the "Fortress Trap" is a transient state or a permanent attractor in evolving social systems.

\begin{credits}
\subsubsection{\ackname} This research was partially supported by the National Science Centre, Poland, grant no. 2022/45/B/ST6/04145 and the Academia Profesorum Iuniorum programme, funded by Wrocław University of Science and Technology.\\
We would like to thank Prof. Katarzyna Sznajd-Weron and Michał Czuba for the consultation regarding the q-Voter model and mABCD, respectively.

\subsubsection{\discintname}
The authors have no competing interests to declare that are relevant to the content of this article.
\end{credits}
\bibliographystyle{splncs04}
\bibliography{sn-bibliography}
\end{document}